\documentclass[12pt]{article}
\usepackage{epsf}
\title{ Spectrum and Regge-trajectories in QCD}
\author{  Yu.A.Simonov\\
 State Research
Center\\Institute of Theoretical and Experimental Physics, \\
Moscow, 117218 Russia}
\newcommand{\beq}{\begin{eqnarray}}
 \newcommand{\eeq}{\end{eqnarray}}
\newcommand{\be}{\begin{equation}}
 \newcommand{\ee}{\end{equation}}

 \def\la{\mathrel{\mathpalette\fun <}}
\def\ga{\mathrel{\mathpalette\fun >}}
\def\fun#1#2{\lower3.6pt\vbox{\baselineskip0pt\lineskip.9pt
\ialign{$\mathsurround=0pt#1\hfil ##\hfil$\crcr#2\crcr\sim\crcr}}}

\newcommand{{\SD}}{\rm SD}

\newcommand{\ver}{\mbox{\boldmath${\rm r}$}}

\newcommand{\vep}{\mbox{\boldmath${\rm p}$}}

\newcommand{\lan}{\langle}
\newcommand{\ran}{\rangle}
\begin{document}
\maketitle
\section{Introduction}

Starting in sixties an active group of physicists under the
guidance of Prof. K.A.Ter-Martirosyan was creating  the theory of
high-energy processes in QCD. From the beginning the key element
of this theory is the notion of Regge trajectories and in
particular of the  pomeron trajectory, which have been introduced
phenomenologically. In this talk I review the problem of spectrum
and Regge trajectories as it can be derived from the
nonperturbative QCD dynamics.

The problem of the QCD spectrum is the central issue in the
nonperturbative QCD and is intimately connected to problems of
confinement and mass generation in QCD. These latter issue make
QCD so much different from QED, in addition explicit form of
Nonperturbative Interaction (NPI) was not known for a long time.
Recently with the introduction of the Field Correlator Method
(FCM) \cite{1,2} the situation has changed favorably, since the
NPI is defined there in a gauge-invariant way and the simple field
correlators, which are sufficient for all dynamical calculations
are known from lattice data \cite{3,4} and analytic results
\cite{5}.

In view of all this one can put the problem of the QCD spectrum in
the most  general framework, as the many-channel problem  of bound
states of quarks, antiquarks and valence gluons with all possible
mixings, the states being stable in the limit $N_c\to \infty$, and
acquiring decay widths when $N_c=3$. It is the purpose of this
talk to describe the QCD spectrum in the Hamiltonian language,
using FCM and the $N_c=\infty$ limit. In addition FCM provides
another convenient limit -- the gluon correlation length $T_g$
tending to zero, while the string tension $\sigma$ is kept
constant. As a result the universal Hamiltonian becomes local and
has a transparent structure for any number of constituents. As
will be seen, it contains only two fixed input parameters:$\sigma$
and $\alpha_s$ (or $\Lambda_{QCD})$ in addition to current quark
masses and is able to predict any meson, baryon, hybrid and
glueball states and their mixings (for a review see \cite{6}).

In doing so one solves the following puzzles: 1) constituent mass
of quarks and gluons  2) meson Regge trajectories with a correct
slope and intercept 3) the explicit notion of the valence gluon 4)
hybrid spectra and Regge trajectories.

Remarkably all calculated spectra are in very good agreement with
lattice data and experiment which gives an additional
justification for FCM. However immediate consequences of the whole
approach are far more reaching. They include a new formulation of
the perturbation theory, the so-called Background Perturbation
theory (BPTh) \cite{7} with $\alpha_s$ saturating at small
Euclidean momenta instead of diverging near the Landau ghost pole.
The whole structure of QCD becomes interconnected with the
spectrum via the quark-hadron duality and hybrids and glueballs
play a very important role in scattering and decay.

\section{Hamiltonian}

There are two possible approaches to incorporating nonperturbative
field correlators  in the quark-antiquark (or $3q$) dynamics. The
first has to deal with the effective nonlocal quark Lagrangian
containing field correlators \cite{8}. From this one obtains
first-order Dirac-type  integro-differential equations for
heavy-light  mesons \cite{8,9}, light mesons \cite{10} and baryons
\cite{10,11}. these equations contain the effect of chiral
symmetry breaking \cite{8} which is directly connected to
confinement.

The second approach is based on the effective Hamiltonian for any
gauge-invariant quark-gluon system. In the limit $T_g\to 0$ this
Hamiltoniam is simple and local, and in most cases when spin
interaction can be considered as a perturbation one obtains
results for the spectra in an analytic form, which is transparent.

For this reason we choose below the second, Hamiltonian approach
\cite{12,13}. We start with the exact Fock-Feynman-Schwinger
Representation for the $q\bar q$ Green's function (for a review
see \cite{14}), taking for simplicity nonzero flavor case
\be
G^{(x,y)}_{q\bar q} =\int^\infty_0 ds_1 \int^\infty_0 ds_2
(z)_{xy}(D\bar z)_{xy}r^{-K_1-K_2}\lan  tr \Gamma_{in} (m_1-\hat
D_1) W_\sigma (C) \Gamma_{out} (m_2-\hat D_2)\ran_A\label{1}\ee
 where $K_i=\int^{s_1}_0 d\tau_i (m_i+\frac14 (\dot
 z_\mu^{(i)})^2), $ $\Gamma_{in,out}=1, \gamma_5,...$ are meson
 vertices, and $W_\sigma(C)$ is the Wilson loop with spin
 insertions, taken along the contour $C$ formed by paths
 $(Dz)_{xy}$ and $(D\bar z)_{xy}$,
 \be
 W_\sigma (C) =P_F P_A\exp (ig \int_CA_\mu dz_\mu)\exp (g\int^{s_1}_0
 \sigma^{(1)}_{\mu\nu} F_{\mu\nu} d\tau_1-g\int^{s_2}_0
 \sigma_{\mu\nu}^{(2)}F_{\mu\nu} d\tau_2).\label{2}\ee
 The last factor in (\ref{2}) defines the spin interaction of
 quark and antiquark. The average $\lan W_\sigma\ran_A$ in
 (\ref{1}) can be computed exactly through field correlators $\lan
 F(1)...F(n)\ran_A$, and keeping only the lowest  one,$\lan F(1)
 F(2)\ran$, which yields according to lattice calculation
 \cite{15} accuracy around 1\%, \cite{16}
 one obtains
$$ \lan  W_\sigma (C)\ran_A \simeq \exp (-\frac12
[\int_{S_{min}}ds_{\mu\nu}(1)
\int_{S_{min}}ds_{\lambda\sigma}(2)+$$ \be+ \sum^2_{i,j=1}
\int^{s_i}_0
 \sigma^{(i)}_{\mu\nu} d\tau_i\int^{s_j}_0
 \sigma_{\lambda\sigma}^{(j)} d\tau_j]\lan F_{\mu\nu}(1) F_{\lambda\sigma}(2)\ran ).
 \label{3}\ee

 The Gaussian correlator $\lan F_{\mu\nu} (1) F_{\lambda\sigma}(2)
 \ran \equiv D_{\mu\nu,\lambda\sigma} (1,2)$ can be rewritten
 identically in terms of two scalar functions $D(x)$ and $D_1(x)$
 \cite{2}, which have been computed on the lattice \cite{3} to
 have the exponential form $D(x) D_1(x) \sim \exp (-|x|/T_g$ with
 the gluon correlation length $T_g\approx 0.2 $ fm.

 Thus the first term in the exponent (\ref{3})
 yields the area law $\lan W_\sigma \cong \ran$ $ \exp (-\sigma S_{\min})$,
 with the string tension $\sigma $ \cite{2}
 \be
 \sigma =\frac12 \int D(x) d^2 x.\label{4} \ee
 We concentrate now on this confining term in (\ref{3})
  and quote the
 result for the spin-dependent term at the end of this section.

 The important point to be discussed now is the character of
 dynamics one gets from (\ref{3}), (\ref{4}). To this end one
 should compare two characteristic lengths (times) $T_g$ and $T_q$
\cite{17}, where $T_q$ is the typical period of quark motion, e.g.
the classical period of motion along the Coulomb orbit (for heavy
quarks) or along the orbit in the linear potential (for light
quarks). In all cases one gets $T_q\ga 1$ fm and hence $T_q\gg
T_g$. Thus one has a local dynamics however relativistic for light
quarks, but in any case the excitation of gluonic vacuum degrees
of freedom can be neglected in the first approximation, so that
the dynamics is of the potential type.

As the next step one introduces the einbein variables $\mu_i$ and
$\nu$; the first one to transform the proper times $s_i, \tau_i$
into the actual (Euclidean) times $t_i\equiv z^{(i)}_4$. One has
\cite{13}
\be
2\mu_i(t_i) =\frac{dt_i}{d\tau_i},~~ \int^\infty_0
ds_i(D^4z^{(i)})_{xy} =const \int
D\mu_i(t_i)(D^3z^{(i)})_{xy}.\label{5}\ee The variable $\nu$
enters in the Gaussian representation of the Nambu-Goto form for
$S_{min}$ and its stationary value $\nu_0$ has the physical
meaning of the energy density along the string. In case of several
strings, as in the baryon case or the hybrid case, each piece of
string has its own parameter $\nu^{(i)}.$

To get rid of the path integration in (\ref{1}) one can go over to
the effective Hamiltonian using the identity
\be
G_{q\bar q} (x,y) =\lan x| \exp (-HT) |y\ran\label{6}\ee where $T$
is the evolution parameter corresponding to the hypersurface
chosen for the Hamiltonian: it is the hyperplane $z_4=const$ in
the c.m. case \cite{13} and $z_+ =const$ in the light-cone case
\cite{18,19}.

The final form of the c.m. Hamiltonian (apart from the spin and
perturbative terms to be discussed later) for  the $q\bar q$ case
is \cite{13,20} $$ H_0=\sum^2_{i=1} \left(
\frac{m^2_i+\vep^2_i}{2\mu_i} +\frac{\mu_i}{2}\right) + \frac{\hat
L^2/r^2}{2[\mu_1(1-\zeta)^2+\mu_2\zeta^2 + \int^1_0 d \beta (\beta
-\zeta)^2\nu (\beta)]}+$$
\be
+\frac{\sigma^2 r^2}{2} \int^1_0 \frac{d\beta}{\nu(\beta)} +
\int^1_0\frac{\nu(\beta)}{2}d\beta.\label{7}\ee

Here $\zeta= (\mu_1+\int_0\beta \nu d\beta)/ (\mu_1+\mu_2+\int^1_0
\beta \nu d \beta)$ and $\mu_i$ and $\nu(\beta)$ are to be found
from the stationary point of the Hamiltonian
\be
\frac{\partial H_0}{\partial\mu_i}|_{\mu_i=\mu_i^{(0)}} =0,~~
\frac{\partial H_0}{\partial\nu}|_{\nu=\nu^{(0)}} =0.\label{8}\ee

Note that $H_0$ contains as input only $m_1, m_2$ and $\sigma$,
where $m_i$ are current masses defined at the scale 1 GeV. The
further analysis is simplified by the observation that for $L=0$
one finds $\nu^{(0)}=\sigma r$ from (\ref{8}) and
$\mu_i=\sqrt{m^2+\vep^2}$, hence $H_0$ becomes the usual
Relativistic Quark Model (RQM) Hamiltonian \cite{21}
\be
H_0(t=0)=\sum_{i=1}\sqrt{m_i^2+\vep^2} +\sigma r. \label{9}\ee For
large $L$ however one can neglect $\mu_i$ as compared to $\nu$ and
one gets
\be
H^2_0\approx 2\pi\sigma \sqrt{L(l+1)},~~ \nu^{(0)}(\beta)= \sqrt
{\frac{8\sigma
l}{\pi}}\frac{1}{\sqrt{1-4(\beta-\frac12)^2}}.\label{10} \ee From
(\ref{9}), (\ref{10}) one can see that $\nu^{(0)}(\beta)$ is
indeed the energy density along the string and $\mu^{(0)}$ is the
c.m. energy of the quark, which plays the role of constituent
quark mass as will be seen below.

To proceed one can use two approximations. First, replace $H_0$ in
(\ref{8}) by its eigenvalue $M_0$, which is accurate within 5\%
\cite{22}. Second, approximate the $L$-dependent term in
(\ref{7}), introducing the correction $\delta H_{str}$  namely
\be
H_0\approx H_R + \Delta H_{str}, ~~ H_R =\sum^2_{i=1} \left(
\frac{\vep^2+m^2_i}{2\mu_i}+\frac{\mu_i}{2}\right) + \sigma
r\label{11}\ee and the mass correction due to $\Delta H_{str} $
for equal mass case is \cite{13}
\be
\Delta_{str} (nL) =\lan \Delta H_{str}\ran= -\frac{16}{3}
\frac{\sigma^2L(L+1)}{M^3_0}\label{12}\ee where $M_0$ is the
eigenvalue of $H_R$; a more accurate approximation is given by
\cite{23} \be \Delta_{str} (nL) =-\frac{2\sigma L(L+1)\lan
1/r\ran}{M_0^2}.\label{13}\ee But $H_0$ is not the whole story,
one should take into account 3 additional terms: spin terms in
(\ref{3}) which produce two types of contributions: self-energy
correction \cite{24} \be H_{self}=\sum^2_{i=1} \frac{\Delta
m^2_q(i)}{2\mu_i},~~ \Delta m^2_q=-\frac{4\sigma}{\pi} \eta
(m_i),~~ \eta(0) \cong 1 \div 0.9\label{14}\ee and spin-dependent
interaction between quark and antiquark $H_{spin}$ \cite{6,25}
which is entirely described by the field correlators $D(x), D_1(x)
$, including also the one-gluon exchange part present in $D_1(x)$.

Finally one should take into account gluon exchange contributions
\cite{7,11}, which can be divided into the Coulomb part $H_{Coul}
=-\frac43\frac{\alpha_s(r)}{r},$ and $H_{rad}$ including
space-like gluon exchanges and perturbative self-energy
corrections (we shall systematically omit these corrections since
they are small for light quarks to be discussed below). In
addition there are gluon contributions which are nondiagonal in
number of gluons $n_g$  and quarks (till now only the sector
$n_g=0$ was considered) and therefore mixing meson states with
hybrids and glueballs \cite{26}. we call these terms $H_{mix}$ and
refer the reader to \cite{26} and the cited there references for
more discussion. Assembling all terms together one has the
following total Hamiltonian in the limit of large $N_c$ and small
$T_g$:
\be
H=H_0+H_{self} +H_{spin} +H_{Coul}+H_{rad} +H_{mix}.\label{15}\ee

We start with $H_0=H_R+H_{string}$. The eigenvalues $M_0$ of $H_R$
can be given with 1\% accuracy by \cite{27}
\be
M^2_0\approx 8\sigma L+4\pi\sigma (n+\frac34)\label{16}\ee where
$n$ is the radial quantum number, $n=0,1,2,...$ Remarkably
$M_0\approx 4\mu_0$, and  for $L=n=0$ one has $\mu_0(0,0)=0.35$
GeV for $\sigma =0.18 $ GeV$^2$, and $\mu_0$ is fast increasing
with growing $n$ and $L$. This fact explains that spin
interactions become unimportant beyond $L=0,1,2$ since they are
proportional to $d\tau_1 d\tau_2\sim \frac{1}{4\mu_1\mu_2}
dt_1dt_2$ (see (\ref{3}) and \cite{25}). Thus constituent mass
(which is actually "constituent energy") $\mu_0$ is  "running".
The   validity of $\mu_0$ as a socially accepted "constituent
mass" is confirmed by its numerical value given above, the spin
splittings of light \cite{28} and heavy-light mesons \cite{29} and
by baryon magnetic moments expressed directly through $\mu_0$, and
being in agreement with experimental values \cite{30}.

The next topic is Regge trajectories in QCD. As it is clear from
(\ref{10}), one has the correct asymptotic Regge slope coinciding
with the string picture, while for small $L, L\leq4$, the
approximation (\ref{11}), (\ref{12}) holds \cite{23} with almost
the same slope. The Regge intercept depends strongly on the term
$H_{self}$ \cite{24}, since (\ref{16}) yields too large value for
$M_0$, e.g. $M_0(0,0)\approx 1.4$ GeV is almost twice the $\rho$
mass. However the self-energy term (\ref{14}) defined
unambigiously through $\sigma$ \cite{24} has a negative sign and
a magnitude which brings the mass back near the experimental
value. Thus one can understand the origin of the large negative
phenomenological constant, which is usually introduced in RQM, but
it is also rewarding that it is not  actually constant, but
depends on $n,L$ via $\mu_0(nL)$ so that the linear  Regge
behaviour is preserved.

To compare with the experiment and  disentangle the contribution
of spin interaction we shall consider the center-of-gravity (cog)
masses for each meson multiplet as in \cite{23}. Then the masses
of all orbital excitations $(n=0)$ can be nicely described by the
linear Regge trajectory, which we call the Regge $L$-trajectory
with experimental parameters \cite{23}
\be
\bar M^2(L) =(1.23\pm 0.02)L+0.37\pm 0.02 ({\rm
GeV}^2)\label{17}\ee or \be L=0.81 \bar M^2(L)-0.30.\label{18}\ee
This is different  from the leading $\rho$-trajectory
\be
J=\alpha'_J M^2(J)+0.48, ~~\alpha'_J=0.88 {\rm GeV}^2,
\alpha_J(0)=0.48\label{19} \ee since  its parameters depend on
spin interactions. Now using (\ref{11}-\ref{16}) one obtains
\cite{23}
\be
L=0.80 \bar M^2(L)-0.34\label{20}\ee which agrees with experiment
(\ref{18}) within 10\%, the accuracy being in accordance with the
estimates of the neglected terms in (\ref{15}), namely $H_{rad}$
and $H_{mix}$. In this way one solves the  problem of Regge
trajectories for orbital excitations in QCD, thus supporting the
foundations o of the wide and fruitful activity undertaken by
Professor K.A.Ter-Martirosyan and his group to describe the
high-energy scattering and production processes in the framework
of the Regge theory. This important contribution was reviewed  in
\cite{31}.

   We now come to the gluon-containing systems, hybrids and
   glueballs. Referring the reader to the original papers
   \cite{32}-\cite{34} one can recapitulate the main results for
   the spectrum. In both cases the total Hamiltonian has the same
   form as in (\ref{15}), however the contribution of corrections
   differs.

   For glueballs it was argued in \cite{34} that $H_0$ (\ref{11})
   has the same form, but with $m_i=0$ and $\sigma\to
   \sigma_{adj}=\frac{9}{4} \sigma$ while $H_{self} =0$ due to
   gauge invariance, and $H_{coul}$ is small due to strong
   cancellation between tree graps and loop corrections \cite{35}.
   Thus glueball masses  are expressed through only $\sigma=0.18$
   GeV$^2$ (fixed by meson Regge trajectories) for the
   center-of-gravity and in addition through $\alpha_s$ for the
   spin splittings. One can see in Table 1 of \cite{34} the theoretical c.o.g.
   mass values computed in \cite{34} in comparison with lattice
   data. The agreement is striking, especially if one takes into
   account that in theoretical calculations there are no
   parameters at all --$\sigma$ was fixed beforehand at the same
   value, as in lattice, and $\alpha_s$ was neglected altogether
   for the reason stated above.

   We now come to the delicate and very important topic of
   glueball trajectories and especially of  the pomeron
   trajectory.  Since the glueball Hamiltonian is basically the
   same as for mesons, one expects that the asymptotic slope of
   all Regge trajectories would be
   \be
   \alpha'_G(M^2\to \infty) \approx
   \frac{1}{2\pi\sigma_{adj}}=\frac49 \alpha'_M.\label{21}\ee
   One expects that the pomeron trajectory is passing through the
   states $2^{++}
$ (2.29 GeV for $\sigma=0.18$ GeV$^2$) and $4^{++}$ (around 3.2
GeV) in which case the pomeron intercept appears too low
$\alpha_P(0)<1$. A possible way out was suggested in \cite{34},
where the intersection of two lowest meson trajectories with the
pomeron trajectory was introduced yielding the correct intercept
$\alpha_p(0)=1.1.\div 1.2$ for reasonable parameters of trajectory
interactions. However this topic is far from clarity, and the BFKL
perturbative results for $\alpha_P(0)$ seem to be not stable with
respect to inclusion to higher orders and/or nonperturbative
effects. It seems that the pomeron trajectory depends on both
perturbative and nonperturbative contributions and their possible
interference, and the   problem was never considered in that
fullness.

 We now coming to the final topic of this talk -hybrids
and their role in hadron dynamics. We start with the hybrid
Hamiltonian and spectrum. This topic in the framework of FCM was
considered in \cite{32,33} The Hamiltonian $H_0$ for hybrid looks
like \cite{6}
\be
H_0^{(hyb)} =\frac{m^2_1}{2\mu_1} + \frac{m^2_2}{2\mu_2} +
\frac{\mu_1+\mu_2+\mu_g}{2} + \frac{\vep^2_\xi+\vep^2_\eta}{2\mu}
+\sigma\sum^2_{i=1} |\ver_g-\ver_i| + H_{str}.\label{22} \ee

Here $\vep_\xi, \vep_\eta$ are Jacobi momenta of the 3-body
system, $H_{self}$ is the same as for meson, while $H_{spin}$ and
$H_{coul}$ have different structure \cite{33}.

The main feature of the present approach based on the BPTh, is
that valence gluon in the hybrid is situated at some point on the
string connecting quark and antiquark, and the gluon creates a
kink on the string so that two pieces of the string  move
independently (however connected at the point of gluon). This
differs strongly from the flux-tube model  where hybrid is
associated with the string excitation  as a whole.

 The difference
between two approaches is especially pronounced in the case of the
hybrid with static quarks \cite{33} where the flux-tube model
predicts for large $R$ the terms $E_n(R)\sim \frac{\pi n}{R}$,
while in FCM there is another branch corresponding to the
longitudinal d.o.f. of gluon $E(R)\sim \frac{const}{R^{1/3}}$
\cite{6,33}, and the lattice results certainly prefer the latter
and contradict the flux-tube  asymptotics. Also at intermediate
interquark distances $R$ the spectrum of FCM approach is in much
better agreement than only another model, see \cite{33} for
details and discussion.

 Results for light and heavy exotic
$1^{-+}$ hybrids are also given in \cite{6} and are in agreement
with lattice calculations. Typically an additional gluon in the
exotic $(L=1)$ state "weights" 1.2$\div$1.5 GeV for light to heavy
quarks, while nonexotic gluon $(L=0)$ brings about 1 GeV to the
mass of the total $q\bar q g$ system.

It is understandable now that hybrids play a very special role in
QCD, namely they describe the excitation of the film --  the
string world sheet -- which is between $q$ and $\bar q$ and in
particular covers the gluon and quark loops appearing in the
$\alpha_s$ renormalization. The "additional gluon" mass
$M_g\approx 1 $ GeV$^2$ cited above enters as a screening mass in
the one-loop running $\alpha_s$ (see \cite{7} \cite{36} for
details and derivations
\be
\alpha_s(Q)=\frac{4\pi}{b_0\ln\left(\frac{Q^2+M^2_g}{\Lambda^2_{QCD}}\right)}\label{23}
\ee and this form (in $r$-space) is in a perfect agreement with a
recent calculation of $\alpha_s$ on the lattice \cite{37}.

Thus one can say that the perturbative QCD in the IR region is
defined by the hybrid physics. A similar conclusion can be drawn
with respect to the DIS diagrams at small to moderate $Q^2, Q^2\la
1 $ GeV$^2$, where familiar ladder diagrams with gluon exchanges
are replaced by (multi)-hybrid Green's functions, and to the
hadron-hadron scattering, where e.g. a sequence of BFKL ladders is
replaced by multihybrid diagrams. The progress in this direction
is highly desirable.

All discussion above refers to the leading large $N_c$ limit,
which has a remarkable accuracy, as it was recently demonstrated
on the lattice \cite{38} and by comparison of calculated and
experimental masses in \cite{23}, \cite{29}, \cite{33}, \cite{34}.
We now turn to the $1/N_c$ effects which as shown in \cite{39}
strongly modify the  masses of radially excited mesons. Typically
those mesons have radii larger than 1.5 fm and  as it was argued
in \cite{39} at the interquark distances  $R\geq 1.4$ fm there
appear sea-quark holes in the confining film connecting $q$ and
$\bar q$ trajectories. It was assumed in \cite{39} that the film
with the virtual holes establishes the quasistationary state which
can be called the "predecay state". It is clear that the effective
string tension decreases due to holes and  the confining potential
is partly screened at $R\geq 1.4$ fm. The masses of radially
excited mesons for $L=0,1,2,3$ and  $n_r=0,1,2,3,4$ have been
calculated in \cite{39} with the help of this potential  and are
shown in Fig.1. One can see a remarkable agreement of theoretical
masses (black dots) with experimental candidates. Another
interesting feature is the almost exact linearity of trajectories
$M^2(n_r, L)$ vs $n_r$. The effect of the mass decrease due to the
sea-quark holes is significant (for  high $n_r$ it is around $0.5$
GeV).

Concluding the talk I would like to stress the  simplicity of the
method (FCM) which  solves at least 4 important problems in the
QCD spectrum: 1) Regge slope 2) Regge intercept  3) the problem of
constituent masses 4) problem of radial Regge trajectories.

The author is indebted for useful discussions to A.B.Kaidalov, Yu.
S. Kalashnikova. During all years of the development of the
present method the author felt a constant interest, a helpful
attention and support from K.A.Ter-Martirosyan, which are very
gratefully acknowledged. The present work was partially supported
by RFBR grants 00-02-17836 and 00-15-96786 and by INTAS grants
00-110 and 00-00366.

\newpage
\begin{figure}[!th]
\epsfysize=15cm
\epsfbox{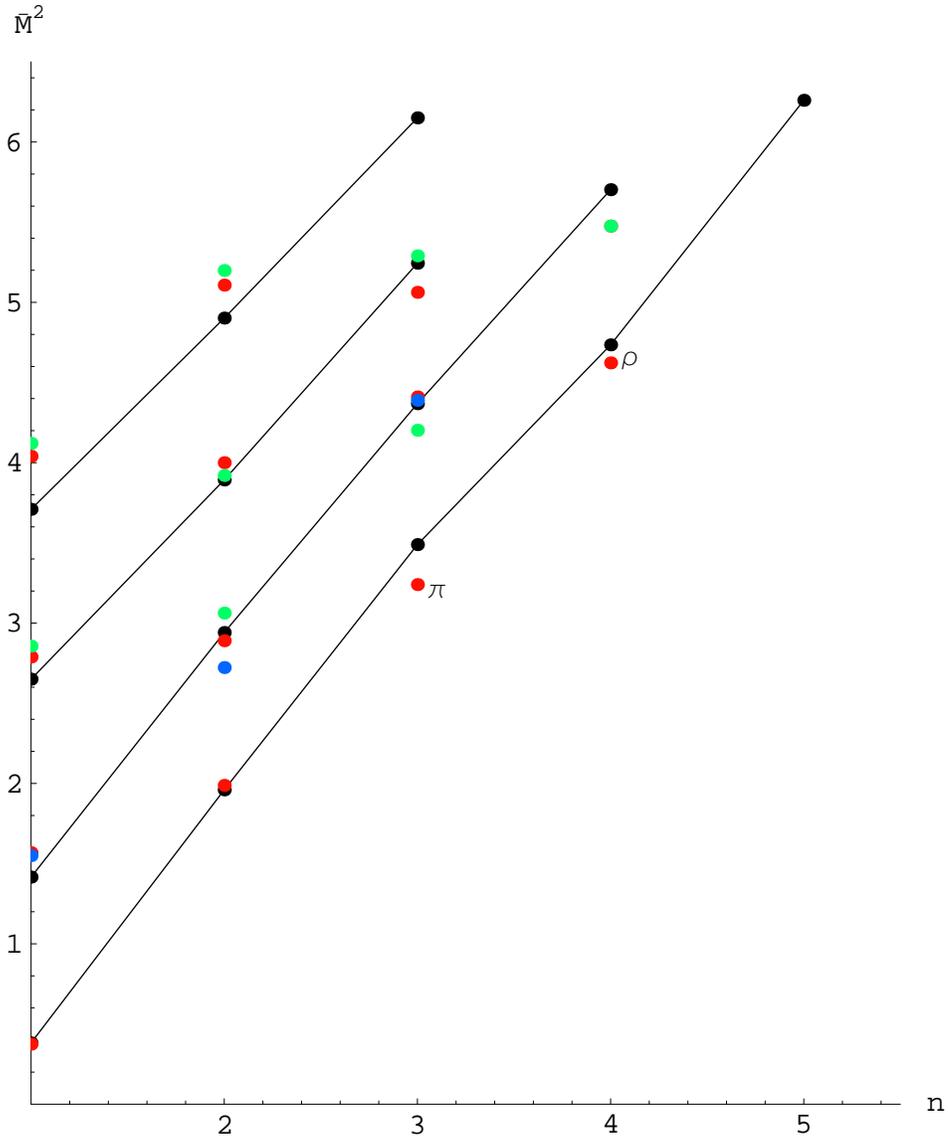}
 \caption{ The spin-averaged
meson mass squared $M^2(L,n_r)$  vs  $n=n_r+1$. Theoretically
calculated masses \cite{39} are depicted by small black dots and
are connected by straight lines.Other points are experimental
candidates.The lines from bottom up correspond to $L=0,1,2,3$
respectively.}
\end{figure}

\end{document}